\documentclass[10pt,conference,anonymous]{IEEEtran}

\usepackage{listings}
\usepackage[skins]{tcolorbox}
\usepackage{fancybox}
\usepackage{changepage}
\usepackage{blindtext}
\usepackage{graphicx}
\usepackage{enumitem}
\usepackage{multirow}
\usepackage{booktabs}
\usepackage{subcaption}
\usepackage{url}
\usepackage{soul,xcolor}
\usepackage{color}
\newcommand{\defj}{\textit{Defects4J}\xspace}
\newcommand{\tbar}{\textit{TBar}\xspace}
\newcommand{\incoder}{InCoder\xspace}

\definecolor{deepblue}{rgb}{0,.2,0.6}
\definecolor{deepgreen}{rgb}{0,0.5,0}
\definecolor{deepchampagne}{rgb}{0.98, 0.84, 0.65}
\definecolor{mintgreen}{rgb}{0.6, 1.0, 0.6}
\definecolor{vividviolet}{rgb}{0.62, 0.0, 1.0}
\definecolor{mangotango}{rgb}{1.0, 0.51, 0.26}

\DeclareFixedFont{\ttb}{T1}{txtt}{bx}{n}{8} % for bold
\DeclareFixedFont{\ttm}{T1}{txtt}{m}{n}{8}  % for normal
\DeclareFixedFont{\ttbb}{T1}{txtt}{bx}{n}{10} % bigger, bold
\DeclareFixedFont{\ttmb}{T1}{txtt}{m}{n}{10}  % bigger, not bold

\definecolor{dkgreen}{rgb}{0,0.5,0}
\definecolor{dkred}{rgb}{0.5,0,0}
\definecolor{gray}{rgb}{0.5,0.5,0.5}

\lstdefinestyle{javastyle} {
language=Java,
basicstyle=\ttfamily\bfseries\footnotesize,
  morekeywords={virtualinvoke},
  keywordstyle=\color{blue},
  ndkeywordstyle=\color{red},
  commentstyle=\color{dkred},
  stringstyle=\color{dkgreen},
  numbers=left,
  breaklines=true,
  numberstyle=\ttfamily\footnotesize\color{gray},
  stepnumber=1,
  numbersep=10pt,
  backgroundcolor=\color{white},
  tabsize=4,
  showspaces=false,
  showstringspaces=false,
  xleftmargin=.23in,
  escapeinside={(*@}{@*)},
}

\newcommand\javainline[1]{{\lstinline!#1!}}
\begin{document}

\title{Revisiting Unnaturalness for Automated Program Repair in the Era of Large Language Models}

\author{
\IEEEauthorblockN{Aidan Z.H. Yang \IEEEauthorrefmark{1}, Sophia Kolak \IEEEauthorrefmark{2}, Vincent J. Hellendoorn \IEEEauthorrefmark{3}, Ruben Martins \IEEEauthorrefmark{4}, Claire Le Goues \IEEEauthorrefmark{5}}
\IEEEauthorblockA{Carnegie Mellon University \\ Pittsburgh United States\\
Email: \IEEEauthorrefmark{1}aidan@cmu.edu, \IEEEauthorrefmark{2}sdkolak@andrew.cmu.edu, \IEEEauthorrefmark{3}vhellendoorn@cmu.edu, \IEEEauthorrefmark{4}rubenm@cs.cmu.edu, \IEEEauthorrefmark{5}clegoues@cs.cmu.edu}}
\maketitle

\begin{abstract}
  Language models have improved by orders of magnitude with the recent emergence
of Transformer-based Large Language Models (LLMs). LLMs have demonstrated their
ability to generate ``natural'' code that is highly similar to code written by
professional developers. One intermediate value an LLM can emit is entropy,
which measures the ``naturalness'' of a token of code. We hypothesize that
entropy can be used to improve the performance of Automated Program Repair (APR)
tasks. While much progress has been made in Automated Program Repair (APR),
fault localization techniques suffer from a lack of diversity in ranking scores,
patch generation tools tend to be inefficient as all tests need to run before
determining if a patch is likely to be correct, and patch ranking often suffers
from the test-suite over-fitting problem. However, using an LLM directly for APR
introduces concerns for training data leakage. In this work, we introduce a
novel way of using the entropy of LLMs in combination with prior APR tools to
improve all stages of APR. By using only the prefix and suffix context of a line
or block of code to describe ``naturalness'', we can use LLMs to localize faults
and rank patches all while eliminating the dependency for test-suites. We show
that entropy is highly complementary with prior fault localization tools. Our
proposed re-ranking method achieves a 50\% Top-5 score improvement over SBFL. We
propose a patch-naturalness measurement, entropy-delta, to improve the
efficiency of template-based repair techniques by ranking plausible patches
before undergoing testing. When using entropy-delta for patch ranking and
classification, our proposed method can rank correct patches more effectively
than state-of-the-art machine learning tools with an 49\% improvement in Top-1.
Our work suggests that LLMs can be an effective addition to compliment prior APR
tasks while minimizing both the test-suite overfitting problem and the LLM data
leakage problem.
\end{abstract}

\IEEEpeerreviewmaketitle

\section{Introduction}
\label{sec:intro}

The problem of software quality has motivated the development of a variety of
techniques for Automatic Program Repair (APR)~\cite{TBar, long_ml_patch,
patch_learning, claire_apr,patch-sim}. At a high level, dynamic APR approaches
use test cases to define a defect to be repaired and functionality to retain,
and to localize the defect to a smaller set of program lines.  
APR techniques generate candidate patches in a variety of ways, such as by
heuristically instantiating pre-defined repair template~\cite{TBar,
kim2022multi}, or by customizing symbolic techniques to synthesize new
code~\cite{patch-sim,angelix}.  

Meanwhile, the recent advances in machine learning and AI, including
but by no means limited to advances in Transformer~\cite{vaswani2017attention}
based language models, have produced orders of magnitude performance improvements over
previous ML techniques for code generation~\cite{Codex,Neox}. ML therefore
affords promising opportunities for program repair~\cite{long_ml_patch,
patch_learning, Infill_repair,Interfix,Prompt_repair} and fault localization~\cite{Llmao}.
The applicability of language models to the repair process makes sense:
these models are trained on large volumes of code in which defects are relatively
rare. Since their training objective encourages next-token prediction,
well-trained language models tend to simultaneously perceive faulty code as unlikely
(or ``unnatural'') and to produce code that is correct, as correct code is more
``natural''~\cite{NAT}. The naturalness of code and unnaturalness of buggy 
code is now a well-established phenomenon~\cite{hindle2016naturalness,NAT}.
However, the bulk of prior research on this topic relied on relatively simple $n$-gram language
models~\cite{ngram}. Compared to present-day LLMs, these models provided a
very poor estimator of code predictability. The ``unnaturalness'' of
buggy lines was therefore mainly useful as an explanatory metric, but showed limited
utility for precisely localizing defects, let alone repairing programs.
The recent advancement of much larger and more sophisticated LLMs have decreased model
perplexities by multiple orders of magnitude.  % Could cite GPT-4 here.
This makes them a much more accurate adjunct both for estimating naturalness and
for fault localization or correct patch identification~\cite{xia2023automated,
yang2023large}.

In this paper, we revisit the idea of (un)naturalness for program repair. The
fundamental idea behind using an LM alone --- even a hypothetically optimal one
--- for repair treats predictability as ultimately equivalent to correctness.
This assumption is specious: LLMs adopt preferences based on a corpus
with respect to training loss that rewards imitation. Beyond the fact that LLMs
necessarily train on buggy code, LLMs generate and score text one token at the
time. Given that, they may well prefer a subtly incorrect implementation spread
across several readable lines over a correct but difficult-to-understand one-line
solution, as the per-token surprisal of the former may be strictly lower than the
latter. Judgement of code correctness requires substantially more context than an
LLM has access to including, but not limited to, test cases test behavior, and
developer intent. Although some of this information could be provided as context,
it will lie outside the training distribution.

This implies that LLMs can only go so far on their own in reasoning about and
fixing buggy code. It moreover motivates the use of traditional tools, which
compress such information, as a complement to LLMs in repair, which has indeed
shown promising recent results for the patch generation stage in
particular~\cite{xia2023automated} (acknowledging the risk of training data
leakage in any such experiment~\cite{balloccu2024leak}). 

% Recent studies has also shown promise in improving LLM and developer code understanding simultaneously in a few-shot setting~\cite{nam2024using}.

We go beyond prior work by interrogating the role of entropy as a complement to
traditional repair at every stage:

\vspace{1ex}
\noindent\textbf{Fault localization (FL).} End-to-end dynamic APR relies on fault
localization to narrow a bug to a smaller set of
source locations. Improving fault localization accuracy is key to improving 
repair efficiency~\cite{OCHIAI, TransferFL}. 
Although FL accuracy is improving, both commonly-used~\cite{abreu2006evaluation}
and state-of-the-art ML-based techniques still suffer from the tendency to
assign the same FL score to large amounts of code.  
For example, we find that
Ochiai SBFL~\cite{abreu2006evaluation} assigns the same suspiciousness score to 1137
lines of code in the dataset \defj (an average of 2.9 ties per bug), and
TransferFL~\cite{TransferFL} assigns the same suspiciousness score to 380 lines
of code in the same dataset (an average of 0.96 ties per bug). 

We show that by incorporating entropy into fault localization, the variance of
suspicious scores increase by 87\% for Ochiai SBFL, and overall accuracy
increases as well (e.g., the Top-5 score improves from 94 to 145).

\vspace{1ex} 
\noindent\textbf{Plausible patch generation.} APR approaches
typically generate multiple potential code changes in search of 
\emph{plausible} patches that cause the program to pass all tests.
Executing tests (and to some extent, compiling programs to be tested) dominates
repair time: the template-based approach \tbar~\cite{TBar} spends about 2\% of
its total time creating patch templates, 6\% generating patches from templates,
and 92\% running tests on generated patches. Regardless of the patch generation method (e.g.,
symbolic techniques~\cite{patch-sim, claire_apr,angelix}, template instantiation~\cite{TBar, kim2022multi}, or
machine learning models~\cite{xia2023automated}) repair \emph{efficiency} is best
approximated in terms of the number of patches that must be evaluated to find a
(good) repair~\cite{efficiency}.

We show that entropy, when used to order candidate patches for evaluation, can
improve the efficiency of generic template-based repair by 24
tested patches per bug, on average.

\vspace{1ex} 
\noindent\textbf{Patch correctness assessment.}
Plausible patches are not always correct, in that they can fail to generalize to
the desired behavior beyond what is tested in the provided test 
suite~\cite{CURE}.  Some recent work aims to address this in the form of a 
post-processing step that identifies (and filters) plausible-but-incorrect
patches, typically by combining program analysis and machine
learning~\cite{Panther, Shibboleth, yang2023large}.  However, techniques to date
are typically trained on the same test suites used for patch generation,
imposing a project-specific training burden (and an expensive one, when dynamic
signals are required), and posing a significant risk of overfitting~\cite{CURE, yang2023large}.  

We show that entropy can rank correct patches 49\% more effectively (in Top-1) than state-of-the-art patch ranker Shibboleth~\cite{Shibboleth}, without using any project-specific training data.

%We incorporate and evaluate these innovations in the context of
%\tbar~\cite{TBar}, the best-performing template-based repair technique for Java
%to date. We do this to mitigate the risk of data leakage in patch generation,
%and to experimentally isolate the utility of entropy at each stage of APR.
%\clg{I feel like that's not well-explained on my part, but it feels like a point
%worth making? But maybe not here? Not sure.}
%\aidan{I think we should move this to RQ2, unless we make this TBar++ RQ4. It would be confusing to say this at the end of the rq3 %patch correctness assessment paragraph.}

In summary, we make the following contributions.
\begin{itemize}
	\item{\textbf{End-to-end entropy APR.} We propose a technique that uses a combination
	of an LLM's inherent ability to detect ``naturalness'' (i.e., entropy) and
	an LLM's generation ability to predict faulty lines, rank untested patches,
	and classify tested but potentially incorrect patches.}
	% \item{\textbf{Fault localization} We propose a technique that uses entropy
	% to re-rank prior state-of-the-art fault localization techniques in the Top-N
	% positions, and show that entropy is highly complementary to prior fault
	% localization tools~\cite{Llmao, TransferFL, abreu2006evaluation}. In
	% particular, we observe that entropy improves the Top-5 score of SBFL from
	% 116 to 145.}
	\item{\textbf{Entropy-delta for efficient template-based patch generation.} 
	We introduce entropy-delta as a
    patch-naturalness measure that can rank patches before running tests. We show
    that entropy-delta can be used to immediately filter out test-failing
    patches, and on average reduce running tests for 24 patches for each bug in
    our dataset. We combine entropy-delta patch ranking with a prior
    template-based program repair technique, \tbar~\cite{TBar}, and release a
    more efficient version of \tbar for future research.}
	% \item{\textbf{Patch disambiguation} We again use entropy-delta to rank
    % plausible (i.e., test passing) patches. We show that entropy-delta can
    % improve the accuracy of prior patch ranking technique
    % PATCH-SIM~\cite{patch-sim} from 0.388 to 0.735. We reduce the test-suite
    % overfitting problem by not training our LLMs on the test-suite used for
    % patch generation.}

	\item{\textbf{Artifact availability}}. Our data, tool, and results are
	available and will be released as open-source.\footnote{\url{https://zenodo.org/records/10851256}}
\end{itemize}

\section{Illustrative Example}
\label{sec:motivating}
%In this section, we discuss a real-world bug and two plausible patches that
%attempts to fix it. We calculate the SBFL score as well as the
%inCoder~\cite{inCoder} entropy values for each code line. We use this example
%to show that the naturalness of code may be an effective tool to improve both
%fault localization and patch disambiguation. 

\lstset{style=javastyle}

\let\origthelstnumber\thelstnumber
\makeatletter

\renewcommand\thelstnumber{%
    \ifnum\value{lstnumber}>0
        \origthelstnumber
    \else
        \ifnum\value{lstnumber}=-1
            \ldots
        \fi
    \fi
}

% for starting line numbers
\newcommand*\startnumber[1]{%
    \setcounter{lstnumber}{\numexpr#1-1\relax}%
}

% for stopping line numbers
\newcommand*\stopnumber{%
    \startnumber{-2}%
}

\begin{figure}[t!]

\begin{subfigure}{\columnwidth}
    \begin{lstlisting}[numbersep=5pt,xleftmargin=21pt,numberstyle=\scriptsize,basicstyle=\footnotesize\ttfamily,firstnumber=4473]
 XYItemRenderer r = getRendererForDataset(d);                      //SBFL=0.52, E=0.48
    (*@ \stopnumber @*) ... (*@ \startnumber{4493} @*) 
- Collection c = getRenderer().getAnnotations();                      //SBFL=0.49, E=1.59 
+ if (r != null) { 
+    Collection c = r.getAnnotations();                       //E=1.34 
(*@ \stopnumber @*) ... (*@ \startnumber{4502} @*)
+ } 
    \end{lstlisting}
    \caption{Chart 4 buggy code and developer fix.}
    \label{fig:motivating-dev}
\end{subfigure}
\begin{subfigure}{\columnwidth}
\begin{lstlisting}[numbersep=5pt,xleftmargin=21pt,numberstyle=\scriptsize,basicstyle=\footnotesize\ttfamily,firstnumber=4493]
- Collection c = getRenderer().getAnnotations();                      //E=1.59 
+ if (r == null) { 
+    return null;         //E=2.77 
+ }
    \end{lstlisting}
    \caption{Chart 4 buggy code and test failing \tbar patch \#1.}
    \label{fig:motivating-fail}
\end{subfigure}
\begin{subfigure}{\columnwidth}
\begin{lstlisting}[numbersep=5pt,xleftmargin=21pt,numberstyle=\scriptsize,basicstyle=\footnotesize\ttfamily,firstnumber=4493]
- Collection c = getRenderer().getAnnotations();                      //E=1.59 
+ if (r == null) {
+    continue;            //E=1.98 
+ }
    \end{lstlisting}
    \caption{Chart 4 buggy code and test passing \tbar patch \#19.}
    \label{fig:motivating-pass}

\end{subfigure}

    \caption{Chart bug 4 from \defj with its developer-written fix, 
    a test-failing patch generated by \tbar, and a test-passing patch generated
    by \tbar. We show the \incoder-produced entropy of code in each patch.}
    \label{fig:motivating}
    \end{figure}

Consider the buggy and fixed versions of (Chart, 4) from
\defj~\cite{just2014defects4j}, shown in Figure~\ref{fig:motivating-dev}. The
original buggy code is missing a null check, which the developer fixed by adding 
\lstinline{if(r != null)} around the
implicated code at line 4493.   

The \tbar~\cite{TBar} template-based program repair technique produces candidate
patches by repeatedly instantiating
applicable templates at program statements, ordered by Ochiai SBFL
suspiciousness score. 
For example, Figure~\ref{fig:motivating-fail} a \tbar-generated patch that does
not cause the tests to pass, and so is discarded, and the search continues.
Given Chart's associated test suite, the Ochiai SBFL approach~\cite{OCHIAI} assigns line 4473
the highest suspicious score in Chart of 0.52; line 4493, a suspiciousness score
of 0.49; and 0.03 to lines 4494 and onward. Using only SBFL for fault localization ranking, the actual faulty line at
line 4493 is ranked as 10th most suspicious. 
This does not prevent \tbar from considering it, but does cost time.  

\tbar can produce patches that pass all tests for this bug, such as the one shown in 
Figure~\ref{fig:motivating-pass}.  In the interest of reasoning about
efficiency, we hold fault localization constant~\cite{efficiency}; given that,
this is the 19th patch attempted.
% This is the 19th patch it will attempt, when
%given perfect fault localization. If using SBFL fault localization, the only test-passing patch occurs in patch \#1076, which uses 2 %hours of compute time on a Intel(R) Xeon(R) 6248R CPU @ 3.00GH. \clg{kind of want to justify this, choice,
%though not sure how to do so briefly, since the choice is reasonable but for
%reasons I struggle to explain succinctly.} \aidan{is this ok?}
% I actually don't know if the perfect FL question is necessary to get into here
% --- TBar will produce this patch eventually even without perfect FL.  
However, although this patch prevents the null pointer exception, it does not
generalize beyond the provided tests to capture the apparent developer intent. Importantly, \tbar \emph{can}
produce the correct patch (from Figure~\ref{fig:motivating-dev}), if configured to execute beyond the  first
test-passing patch found --- it is the 70th patch attempted, but only the second
that passes all tests.\footnote{Using SBFL fault localization, \tbar
produces these patches at 1076 and 1127, respectively.} 

LLM-based entropy provides useful clues, here, however.
First, consider fault location: we use
\incoder~\cite{inCoder},\footnote{When prompted
with the code and asked to fix the bug directly, \incoder does not produce a test-passing patch in few-shot setting. Note that GPT4 fixes the
bug correctly, and reports the git commit associated with the fix, implicating
data leakage.} to measure the entropy of every line in this file. Rank-ordering
them, line 4473 is ranked 8th-most-surprising.  This is better than the SBFL
technique on face.  However, their real utility appears to lie in
combination: re-ranking the lines receiving the Top-10 SBFL suspicious scores by
\incoder entropy values, puts line 4493 at rank 2. 
We investigate how entropy in conjunction with SBFL performs for fault localization across multiple bugs
and projects, as well as how different LLMs affect its performance.

We can also measure the naturalness of generated
patches, such as by calculating the change in entropy, which we call
entropy-delta ($\triangle E$),
between the original buggy line of code and the proposed patches. The $\triangle
E$ for the test-failing patch is $-0.39$; for the test-passing but
still-incorrect patch is $-1.18$; and for the correct patch is $0.25$.

The entropy-delta scores do not perfectly predict behavior (note that the
test-failing patch has a higher score than the test-passing-but-incorrect
patch), but it still suggests: 
\begin{enumerate}
    \item Entropy-delta can improve efficiency by suggesting the order to test
    patches.  Test execution time is the dominant cost in program repair.  By using entropy to rank potential patches
    \emph{before} testing them, to suggest the order in which to do so, both
    test-passing patches can be found within 6 attempts (improving on 19 to the
    first test-passing in the default mode, and 90 to find the second, correct
    patch).
    \item Entropy-delta can potentially help disambiguate
    plausible-but-incorrect from genuinely correct patches.
\end{enumerate} 

%developers typically inspect onl the Top-10 suggested elements by repair
%techniques~\cite{kochhar2016practitioners}.
%CLG: I would love to include that reference somewhere, but not sure it still
%fits here. 

We evaluate these relationships in detail in the rest of this work, showing how
entropy can usefully complement traditional approaches to automatic localization
and transformation in the context of program repair.

\section{Approach}
\label{sec:method}
\begin{figure*}[t]
\centering
\includegraphics[width=\textwidth]{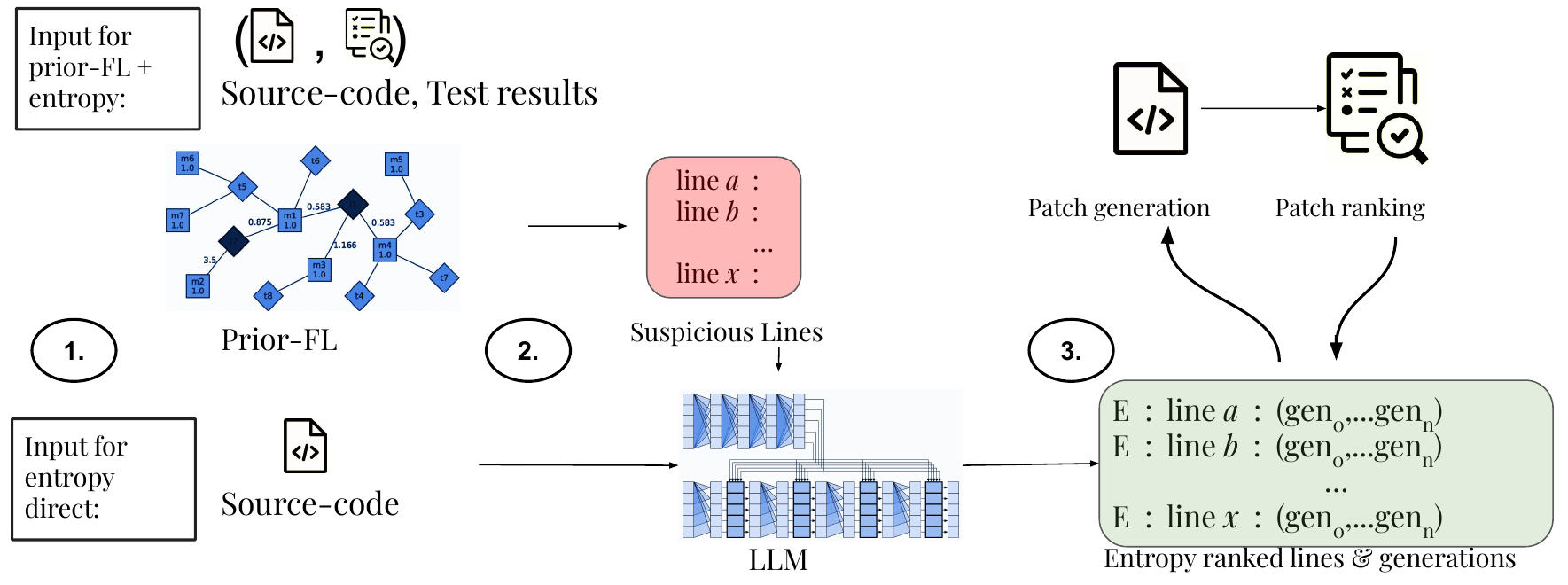}
\caption{Fault localization pipeline using entropy. (1) We take a prior-FL suspicious score list, (2) query each code-line for entropy values, and (3) re-rank the list using LLM entropy scores.}
\label{fig:fl-method}
\end{figure*}

% \begin{figure}[t] \centering
% \includegraphics[width=.48\textwidth]{images/generation_query.pdf} \caption{An
% example of entropy-delta query from a code-line deletion patch.}
% \label{fig:deletion-method} \end{figure*}

% \begin{figure}[t] \centering
% \includegraphics[width=.48\textwidth]{images/entropy_query.pdf} \caption{An
% example of entropy-delta query from a code-line replacement patch.}
% \label{fig:replacement-method} \end{figure}

\begin{figure}
   \begin{subfigure}{\columnwidth}
   \includegraphics[width=\textwidth]{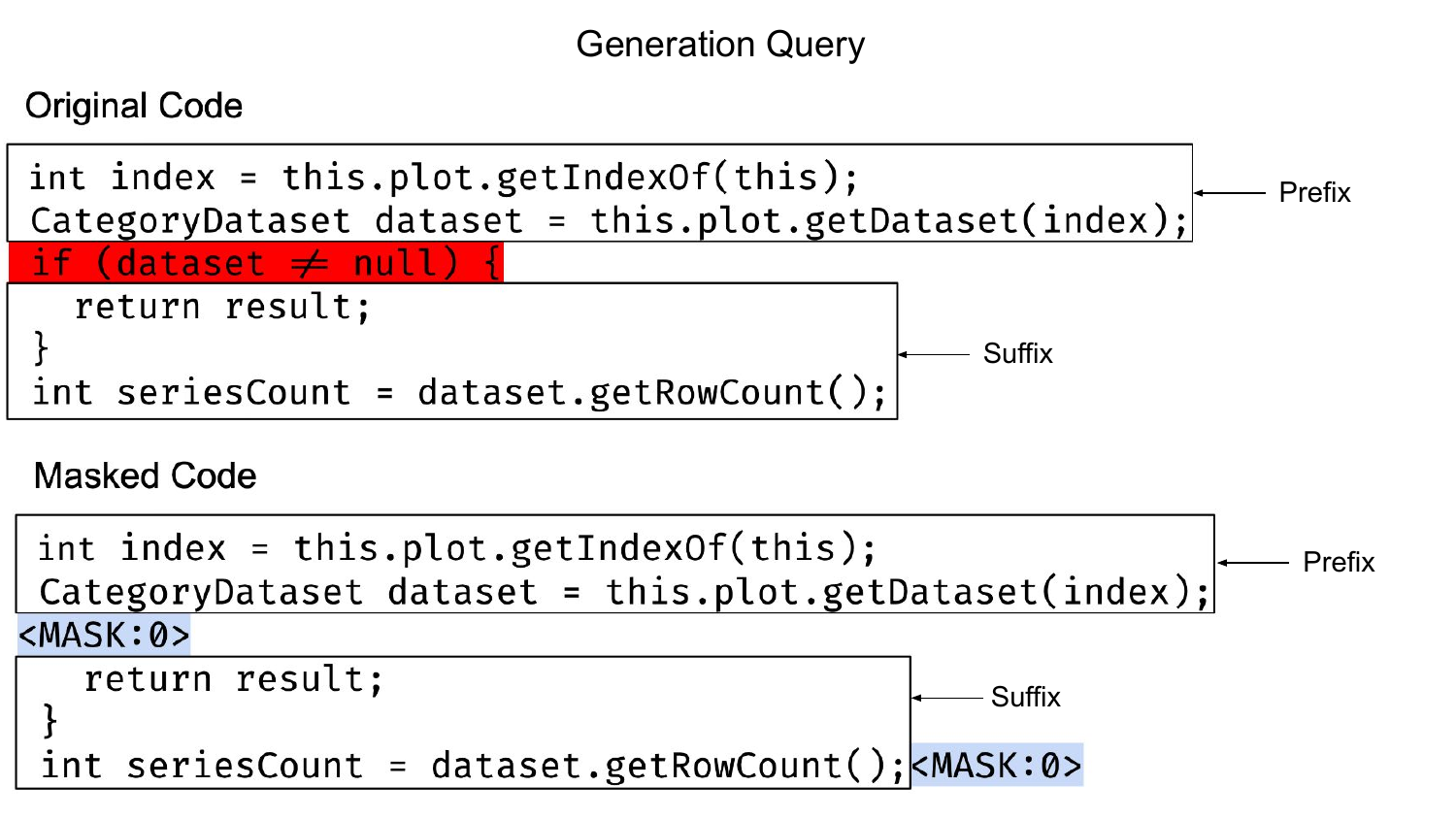}
   \caption{An example of entropy-delta query from a code-line deletion patch. The entropy-delta value of the deleted line is the difference between the original line and a blank line.}
   \label{fig:deletion-method}
   \end{subfigure}

   \begin{subfigure}{\columnwidth}
      \includegraphics[width=\textwidth]{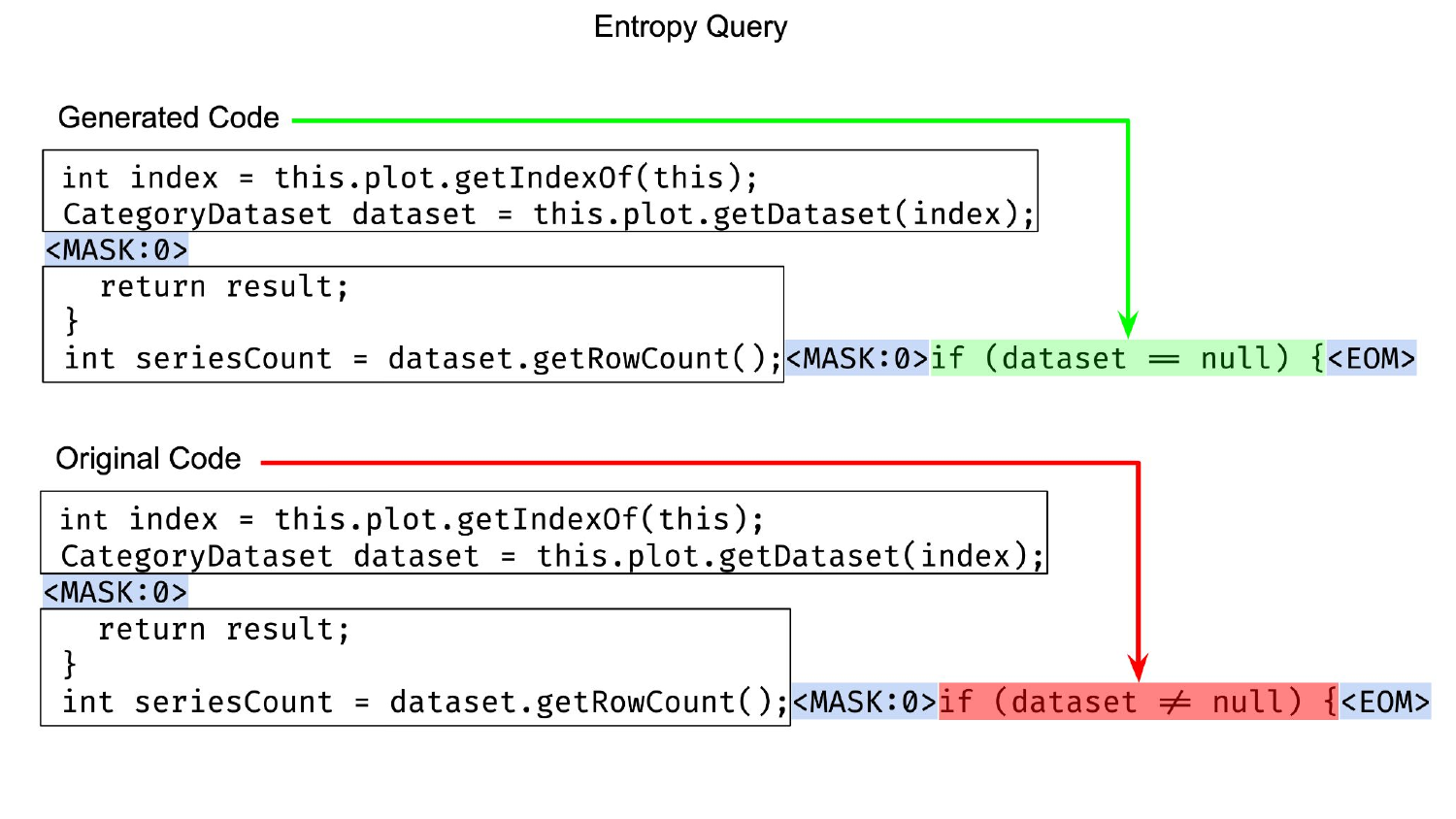}
      \caption{An example of entropy-delta query from a code-line replacement patch. The entropy-delta value of the replaced line is the difference between the original line and the replacement line.}
      \label{fig:replacement-method}
   \end{subfigure}
   \caption{Example entropy-delta queries from an LLM. The MASK tokens enable models to learn the contextual relationships between tokens and make entropy predictions for missing, new, or replacement tokens. The EOM token is a special token that indicates the end of a mask.}
\end{figure}

We ask and answer the following three research
questions about the utility of LLM-entropy for APR.
\begin{itemize}
    \item \textbf{RQ1: How can entropy improve fault localization?} We perform an
   empirical evaluation of prior state-of-the-art fault localization tools and
   observe whether and how they benefit from the use of entropy scores. 

    \item \textbf{RQ2: How can entropy improve patch generation efficiency?} To
    measure how entropy can be used for patch generation efficiency, we use
    it to rank proposed patches generated by an APR technique before
    running tests.
    \item \textbf{RQ3: How well does entropy-deltas identify correct patches?} We
    investigate if entropy-delta can differentiate plausible patches (patches
    that passes all tests) and correct patches (patches that correctly fix the
    bug).

\end{itemize}

This section describes how we use entropy for fault localization (Section~\ref{sec:entropy-fl});
 our development of entropy delta for evaluating patches
 (Section~\ref{sec:entropy-delta}); and our modifications to \tbar to enable our
 study of improved patch efficiency (Section~\ref{sec:entropy-tbar}).  The next
 section describes datasets and metrics.

\subsection{Entropy for fault localization}
\label{sec:entropy-fl}

We integrate raw entropy scores into prior fault localization techniques.  
Figure~\ref{fig:fl-method} overviews the approach.  We take suspiciousness scores
provided by a given prior FL tool for a given file. For consistency with prior
studies of fault localization, we focus on top-N identified lines (as developers
do not typically inspect more than 5
candidates~\cite{kochhar2016practitioners}).  We choose 6 and 10 
for N, and name the two approaches 6-filter and 10-filter, as these are two
quantities near the Top-5 that can 
still significantly impact Top-5 scores. 

We then query an LLM for entropy scores for each line of code in that file. We
tokenize the entire file, iteratively masking each line, and querying the model
for each line's entropy. We used a sliding context window with 2048 tokens
(i.e., the maximum attention window of our smallest selected LLM) surrounding
the mask as suffix and prefix context. This allows us to assign entropy-based
scores for all code in a file, even those longer than a given LLM's context
window. 

We then re-rank the suspiciousness code identified by SBFL by entropy score, and
validate the ranked list according to the actual fault location in the dataset.

We incorporate entropy into three previous FL techniques: SBFL using the Ochiai
formula~\cite{OCHIAI}, TransferFL~\cite{TransferFL} and LLMAO~\cite{Llmao}.
Ochiai is a common formula in SBFL used in traditional APR practices (e.g.,
\tbar). TransferFL and LLMAO are the current state-of-the-art FL techniques using transfer-learning and LLMs, respectively.

\subsection{Entropy-Delta}
\label{sec:entropy-delta}

To evaluate patch naturalness, which we use in both patch prioritization during
generation/evaluation and patch correctness prediction, we introduce the concept of an
``entropy-delta''. Entropy delta describes how code replacement changes the naturalness
of a block of code. Figure~\ref{fig:deletion-method} and
Figure~\ref{fig:replacement-method} give examples for our usage of entropy-delta
for assigning a ranking score for patches. Figure~\ref{fig:deletion-method}
shows the process of masking out a deleted line of code and querying the LLM for
the change in entropy using that mask (i.e., the change in entropy without the original line). Figure~\ref{fig:replacement-method} shows
the process of querying the LLM for the change in entropy if the tokens of the
original line of code is replaced with new tokens of patch code. If the patch is
an insertion of a blank new line, we query the entropy-delta between the ``newline'' token and the original line of code.
For the case of an insertion, we measure the entropy-delta between the new code line and the original blank line.

An entropy-delta is simply the change in entropy before and after a line in code
is replaced. For instance, if the line's original entropy is 1.0, and the
replacement line's entropy is 0.5, then the line has an entropy-delta of +0.5,
as in, replacing that line lowered entropy by 0.5. A significant reduction in
entropy (large, positive entropy-delta) means that the replacement code lowered
the entropy, implying both that the original statement may have been buggy and
that the patch is more natural for that region of code. A large, negative
entropy-delta means that the replacement code increased entropy, meaning that
the patch is less natural at that location. An entropy-delta of 0 means that the
patch has the exact same naturalness as the original code.

\subsection{Modified TBar}
\label{sec:entropy-tbar}

Our patch efficiency experiments ask how entropy can speed up patch generation
and evaluation.  We evaluate it in context of \tbar~\cite{TBar}, the best-performing
template-based program repair technique in the existing literature.  We avoid
using ML-based APR techniques (even though some may outperform
\tbar~\cite{xia2023automated, TransferFL, Dear}) because our goal is a controlled evaluation of entropy without learned patterns from the test suite.  Evaluating based on a technique that otherwise also relies on trained ML
models fails to isolate the effect of entropy per se.

\tbar is a template-based patch generation technique
integrated with \defj V1.2. Our experiments require several modifications to the
codebase.  First, we enable \tbar to continue seeking patches after the first
test-patching patch is found.  Second, we enable \tbar to generate patches, or
evaluate them in a customized order (such as one provided by an entropy-delta ranking).
Our \tbar extension also includes some refactoring for
modifiability/extensibility, as well as a more accurate patch caching mechanism
(caching the patched source code, rather than the patch alone).  We provide the
modified code with our replication package.

\section{Datasets and metrics}
\label{sec:setup}
In this section, we describe the models we use for entropy
(Section~\ref{sec:llms}), the bug and patch datasets considered
(Section~\ref{sec:dataset}), as well as evaluation metrics (Section~\ref{sec:metrics}).

\subsection{LLMs}
\label{sec:llms}

We used \incoder~\cite{inCoder}, Starcoder~\cite{Starcoder}, and Code-Llama2
\cite{Llama}. The three LLMs were trained on open-source code and are capable of
infilling with bidirectional context. The \incoder model~\cite{inCoder} is a
large-scale decoder-only Transformer model with 6.7 billion parameters. The
model was trained on a dataset of code from public open-source repositories on
GitHub and GitLab, as well as from StackOverflow. 
% The dataset includes code
% written in various programming languages, with a focus on Python and JavaScript.
%The model was also trained on code written in 28 other programming languages.
\incoder was primarily trained for code infilling, which involves the insertion
of missing code snippets in existing code, using a causal-masked objective
during training. However, its versatility enables it to be utilized in a variety
of software engineering tasks, including automatic program repair. Starcoder and
Llama-2 were trained with a similar autoregressive plus causal-mask objective as
\incoder. Starcoder was trained with 15.5 billion parameters. Code-Llama2 have
three versions available: 7B, 13B and 34B. We choose the 7B version as it is
the closest in size to the other two models. Although the three LLMs were
not specifically trained for repair, their large architectures and training
objectives could imply that their entropy values on a particular region of code
could suggest code naturalness. For all experiments, we set the LLM temperature
to 0.5.

\subsection{Dataset}
\label{sec:dataset}

\begin{table}[t]
    \centering
  \caption{\small \defj bugs with at least one patch passing tests (RQ2 - efficiency), and a developer fix (RQ3 - patch correctness).}
  \begin{tabular}{l|rr|rr}
  \toprule
  & \multicolumn{2}{c}{\textbf{\defj V1.2 \#bugs}}  & \multicolumn{2}{c}{\textbf{\defj V2.0 \#bugs}}  \\
  & \multicolumn{2}{c}{Patch efficiency (RQ2)}  & \multicolumn{2}{c}{Patch
  correctness (RQ3)}  \\
  & Incl. & Total & Incl. & Total \\
  \midrule
  Chart  & 11  & 26 & 19  & 26\\
  Closure  & 19 & 133&  64 & 174\\
  Lang  & 14 &65& 35 & 64\\
  Math  & 21 & 106 & 67  & 106\\
  Mockito  & 3 & 38&  1  &38\\
  Time  & 4 & 27  & 11 & 26\\
  \midrule
  Total & 72 & 395 & 197 & 434 \\
  \bottomrule
  \end{tabular}
  \label{table:datasets}
  \end{table}

We use the \defj~\cite{just2014defects4j} dataset as the basis of our
experiments.  \defj is a well-established set of documented historical bugs in Java
programs with associated tests and developer patches. It is commonly used in
APR, testing, and fault localization research. However, each research question requires a different subset of the
data. Table \ref{table:datasets} shows the number of bugs in each project that
have at least one patch passing tests (for analyzing patch efficiency) and a
developer fix (for analyzing patch correctness) along with plausible but
incorrect patches. In total, we analyze 72 bugs from \defj V1.2 for patch
efficiency and 197 bugs from \defj V2.0 for patch correctness.

We used \defj V1.2 for the fault localization and patch generation experiments. 
We do this because off-the-shelf TBar, as well as prior fault localization tools' replication packages, are all only compatible with \defj V1.2. The fault localization experiments consider all 395 bugs in \defj V1.2.  
We choose not to use \defj V2.0 for fault localization because prior tools' replication packages are only compatible with \defj V1.2. 

For patch generation, the goal is to evaluate the degree to which entropy can improve repair efficiency; we therefore focus on the subset of \defj V1.2 bugs on which vanilla \tbar succeeds at least once.

% For RQ3, we are interested in patch disambiguation on patches that passes all tests. 
% To make a fair comparison against state-of-the-art patch ranking and patch classification tools, we use the curated dataset from prior tools' replication packages directly, which is built on top of the updated \defj V2.0 dataset. 

For patch correctness ranking, we use curated datasets from prior tools'
replication packages directly, namely, Shibboleth~\cite{Shibboleth} and
Panther~\cite{Panther}. Shibboleth and Panther are both tools that leverage
static and dynamic heuristics from both test and source code to rank and
classify plausible patches, built on top of the updated \defj V2.0 dataset. We
use a dataset of 1,290 plausible patches on \defj V2.0 curated by Ghanbari et
al.~\cite{Shibboleth}. For patch classification, we use a dataset of 2,147
plausible patches on \defj V2.0 curated by Tian et al.~\cite{Panther}. The
patches from Tian et al.~\cite{Panther} were generated by seven different APR
techniques. Each bug in the data set has one correct patch and several plausible
(i.e., test passing) but incorrect ones. We calculate the change in entropy
between the section of code in the original (buggy) file and the patched
version. Note that both datasets only contain patches in projects Chart, Closure, Lang, Math, Mockito, and Time (6/17 of \defj V2.0's total projects), to compare with prior work built on \defj V1.2. Instead of the total number of bugs 835 in \defj V2.0, we only consider the 434 bugs in the 6 projects included by Shibboleth~\cite{Shibboleth} and
Panther~\cite{Panther} (shown in Table \ref{table:datasets}).

\subsection{Metrics}
\label{sec:metrics}

\noindent\textbf{Fault localization and patch ranking.} We measure the effectiveness of both fault localization ranking and patch
ranking by counting the number of correct faults or patches that appear in the
Top-N position. The Top-N measure has been widely used in APR
research~\cite{wong2016survey}. Existing studies~\cite{kochhar2016practitioners}
showed that over 70\% of developers inspect only the Top-5 suggested elements.
We use Top-5, Top-3, and Top-1 for fault localization ranking. We only use Top-2
and Top-3 for patch ranking following Ghanbari et al.~\cite{Shibboleth}, as some
bugs in our dataset only have 2 plausible patches available.

\noindent\textbf{Patch generation efficiency.} We measure the effect of
reranking generated potential patches in terms of the number of patch
evaluations saved by doing so.  Patch evaluations are established as a
hardware- and program-independent measure for APR efficiency~\cite{efficiency},
and a proxy for compute time. 

\noindent\textbf{Patch correctness.}
For patch classification tasks, we convert entropy-delta values into binary labels. We
label patches with a positive entropy-delta as ``more natural'' (i.e., more
likely to be correct), and patches with a negative entropy-delta ``less
natural'' (i.e., less likely to be correct). To measure entropy's ability to
isolate correct and incorrect patches, we use +recall and -recall. +Recall
measures to what extent correct patches are identified, while -recall measures to
what extent incorrect patches are filtered out.  We use accuracy, precision, and
F1 scores to assess classification effectiveness over the entire dataset. 

%In all definitions below, TN is
%true negative, TP is true positive, FP is false positive, and FN is false
%negative.

%\begin{equation}+Recall = \frac{TP}{TP + FN}\end{equation}
%\begin{equation}-Recall = \frac{TN} {TN + FP}\end{equation}

%To have an understanding of the classification effectiveness for the entire
%dataset, we use accuracy, precision, and F1, which are defined as follows:
%\begin{equation}Accuracy = \frac{TP + TN} {TP + TN + FP + FN}\end{equation}

%\begin{equation}Precision = \frac{TP}{(TP + FP)}\end{equation}

%\begin{equation}F1 = \frac{TP} {TP + 0.5(TP + FN)}\end{equation}

\section{Results}
\label{sec:results}

% \begin{figure*} \centering \subfigure[\todo{14 pt font, add medians to LHS
%   plot. Remove RHS
%   plot.}]{\includegraphics[scale=0.38]{images/methods.png}}\quad
%   \subfigure[random caption
%   2]{\includegraphics[scale=0.38]{images/top3-wb.png}} \end{figure*}

\begin{table}[t]
  \centering
\caption{\small Top-N scores on 395 bugs from \defj V1.2.0 from 3 prior tools and re-ranking with entropy from three pre-trained LLMs: \incoder (6B), LLAMA-2 (7B), and Starcoder (15.5B)}
\begin{tabular}{l|p{1cm}lrrr}

\toprule
 \textbf{FL type} &  \textbf{re-rank Filter} & \textbf{Technique} &
 \textbf{Top-1} & \textbf{Top-3} & \textbf{Top-5}\\
\midrule
\multirow{1}{*}{Entropy}
&& entropy-Llama2                         & 5  & 20  & 41   \\
&& entropy-Starcoder                      & 9  & 35  & 55   \\
&& entropy-\incoder                      & \textbf{38}  & \textbf{90}  &
\textbf{116}   \\

\midrule
\multirow{3}{*}{SBFL}
&& SBFL                         & 24  & 61  & 94   \\
\cmidrule{3-6}
&\multirow{2}{*}{10-filter} & entropy-Llama2           & 15  & 37  & 84   \\
&& entropy-Starcoder       & 15  & 40  & 88   \\
&& entropy-\incoder         & 50  & 98  & 133   \\
\cmidrule{2-6}
&\multirow{2}{*}{6-filter} & entropy-Llama2         & 25  & 84  & \textbf{145}
\\
&& entropy-Starcoder       & 28  & 86  & 144   \\
&& entropy-\incoder        & \textbf{55}  & \textbf{117}  & 141   \\

\midrule
\multirow{3}{*}{TransferFL}
&& TransferFL                 & \textbf{69}  & 126  & 145   \\
\cmidrule{3-6}
&\multirow{2}{*}{10-filter} & entropy-Llama2             & 33  & 82  & 105   \\
&& entropy-Starcoder         & 39  & 92  & 126   \\
&& entropy-\incoder           & 49  & 114  & 144  \\
\cmidrule{2-6}
&\multirow{2}{*}{6-filter} & entropy-Llama2             & 38  & 131  &
\textbf{184}   \\
&& entropy-Starcoder         & 44  & 138  & 178   \\
&& entropy-\incoder           & 57  & \textbf{156}  & 182   \\

\midrule
\multirow{3}{*}{LLMAO}
&& LLMAO                        & \textbf{87}  & 134  & \textbf{149}   \\
\cmidrule{3-6}
&\multirow{2}{*}{10-filter} & entropy-Llama2                & 38  & 131  & 145
\\
&& entropy-Starcoder            & 45  & 107  & 144   \\
&& entropy-\incoder              & 77  & 131  & 146   \\
\cmidrule{2-6}
&\multirow{2}{*}{6-filter} & entropy-Llama2                & 49  & 121  & 143
\\
&& entropy-Starcoder            & 36  & 84  & 151   \\
&& entropy-\incoder              & 81  & \textbf{142}  & \textbf{151}   \\

\bottomrule
\end{tabular}
\label{table:fl_topn}
\end{table}
  
In this section, we present results on the performance of entropy and entropy-delta on our three
research questions: 

\begin{enumerate}[leftmargin=3em]
    \itemsep0em 
    \item[\textbf{RQ1:}] Can entropy improve fault localization?
    \item[\textbf{RQ2:}] Can entropy improve patch generation efficiency?
    \item[\textbf{RQ3:}] How well does entropy-deltas identify correct patches?
\end{enumerate}

\subsection*{RQ1: Can entropy improve fault localization?}
In this research question, we compared 24 different configurations for fault
localization. Our analysis aims to determine the most effective approach for
identifying the buggy statement in a series of one line bugs. We first measure
entropy directly for fault localization with our three selected LLMs:
Code-Llama2, Starcoder, and \incoder. We then measure the fault localization
accuracy of three prior fault localization tools: SBFL~\cite{OCHIAI}, TransferFL~\cite{TransferFL}, and LLMAO~\cite{Llmao}.
Finally, we use entropy to re-rank prior fault localization tools and observe
that entropy re-ranking largely improves prior tools. Table \ref{table:fl_topn}
shows the Top-N scores (N = 1,3,5) on all configurations of our experiment. We
observe that the entropy of \incoder, the smallest LLM in our lineup, is
the most effective for fault localization. This is consistent with results
from Xia et al.~\cite{xia2023automated}, who found
that \incoder, trained with an objective of predicting missing code from a
bidirectional context, is more effective at program repair tasks than larger
but purely causal generative LLMs.

\noindent\textbf{SBFL.} From Table \ref{table:fl_topn}, we observe an overall
decrease in Top-N scores using either Code-Llama2 or Starcoder entropy with a
10-filter. However, all Top-N scores improve with the 6-filter. In particular,
the Top-3 score of 84 for Llama2 entropy improves upon SBFL by 38\%, and the
Top-3 score of 86 for Starcoder entropy improves upon SBFL by 41\%. Using
\incoder entropy to re-rank SBFL  shows substantial improvements across all Top-N
and the two types of filters. \incoder entropy-SBFL with a 10-filter achieves a
Top-1 score of 50 (108\% improvement), and 5-filter achieves a Top-1 score of 55
(129\% improvement). Similarly, the Top-3 and Top-5 scores improve by 61\% and
41\%, respectively for \incoder entropy-SBFL with a 10-filter. The Top-3 and
Top-5 scores improve by 92\% and 50\% respectively for the 10-filter.

\noindent\textbf{TransferFL.} As seen in Table \ref{table:fl_topn}, we observe
an improvement from entropy on TransferFL's Top-3 and Top-5 scores using a
6-filter. In particular, 6-filter \incoder-entropy with TransferFL Top-3 is 156
(24\% improvement), and 6-filter Llama2-entropy with TransferFL Top-5 is 184
(27\% improvement). However, 6-filter \incoder-entropy with TransferFL yields a
Top-1 score of 57, which is a 17\% decrease in performance than TransferFL by
itself. As compared to state-of-the-art machine learning based FL techniques, we
observe that entropy scores perform worse on Top-1.
% We notice that the correct fault location is not in the Top-6, so
% filtering for only the Top-6 and re-ranking only decreases fault localization
% performance. 

%\ruben{I do not understand what you are trying to say here. How is the correct
% fault location not in the Top-6? Looking at the other columns, it found 182 bugs
% on Top-5, so the re-ranking has the potential to put the lines on Top-1, it just
% does not do it. Are the entropy scores too close?}
% \aidan{I deleted that but still don't know what kind of discussion I can say, looks like entropy is just worse.}

\noindent\textbf{LLMAO.} Similar to the results of TransferFL, re-ranking with
entropy only improves fault localization results using the 6-filter.
Furthermore, only entropy calculated using \incoder shows an improvement over
LLMAO alone for Top-3 and Top-5, with a 8\% and 1\% improvement, respectively.
Since LLMAO is already an LLM based FL tool, LLM entropy re-ranking shows
marginal improvements as compared to prior non-LLM based FL tools. LLMAO
finetunes on CodeGen 16B~\cite{nijkamp2022codegen}, which is a larger LLM than
our three chosen LLMs.

Our results indicated that SBFL benefits the most with \incoder's entropy
re-ranking. Since SBFL has the most amount of tied suspicious scores (2.9 ties
per bug on average), the additional suspiciousness from entropy values helps to
break ties. 
% We find that the variance on all entropy suspicious scores across
% lines of code in \defj is 0.071 (87\% higher than that of SBFL suspicious
% scores). \ruben{We probably want to remove the previous line. What does it mean to have a variance of 0.071 on the entropy scores for suspicious? How is the variance computed and how do you get the number 87\% higher than that of SBFL? What is the goal of this sentence?} In contrast, TransferFL has an average of 0.96 ties per bug, and LLMAO
% has no ties for suspicious scores. 
TransferFL and LLMAO benefit from entropy
re-ranking mostly when using a 6-filter. These findings suggest that
incorporating entropy as a heuristic in fault localization can improve the
accuracy of identifying the buggy statement, particularly when used in
conjunction with SBFL. 

\begin{tcolorbox}
  [colback=white,colframe=black,arc=0pt,boxrule=0.5pt,title=RQ1
   Summary,boxsep=2pt,left=1pt,right=1pt,top=1pt,bottom=1pt,fonttitle=\bfseries]
   We leverage entropy for fault localization in \defj programs and show that,
   while entropy alone is only somewhat useful for finding defective lines, the
   measure is highly complementary when combined with prior fault localization
   tools, which highlights the importance of combining LLM-based methods with
   techniques from prior APR approaches.
  \end{tcolorbox}

\subsection*{RQ2: Can entropy improve patch generation efficiency?}

\begin{table}[t]
  \centering
\caption{\small entropy-delta ranking scores of 72 plausible patches generated by \tbar per \defj project. The mean rank decrease is 24 and the median is 5.}
\begin{tabular}{l|rr}
\toprule
\textbf{Project}  & \textbf{Improves ranking}& \textbf{Lowers ranking}   \\
\midrule
Chart  & 11 &  2 \\
Closure  & 15 &  4 \\
Lang  & 11 &  2 \\
Math  & 16 &  3 \\
Mockito  & 1 &  0 \\
Time  & 3 &  1 \\
Total & 60 & 12 \\
\bottomrule
\end{tabular}
\label{table:patch_efficiency}
\end{table}

\begin{figure}[t]
  \centering
  \includegraphics[width=0.5\textwidth]{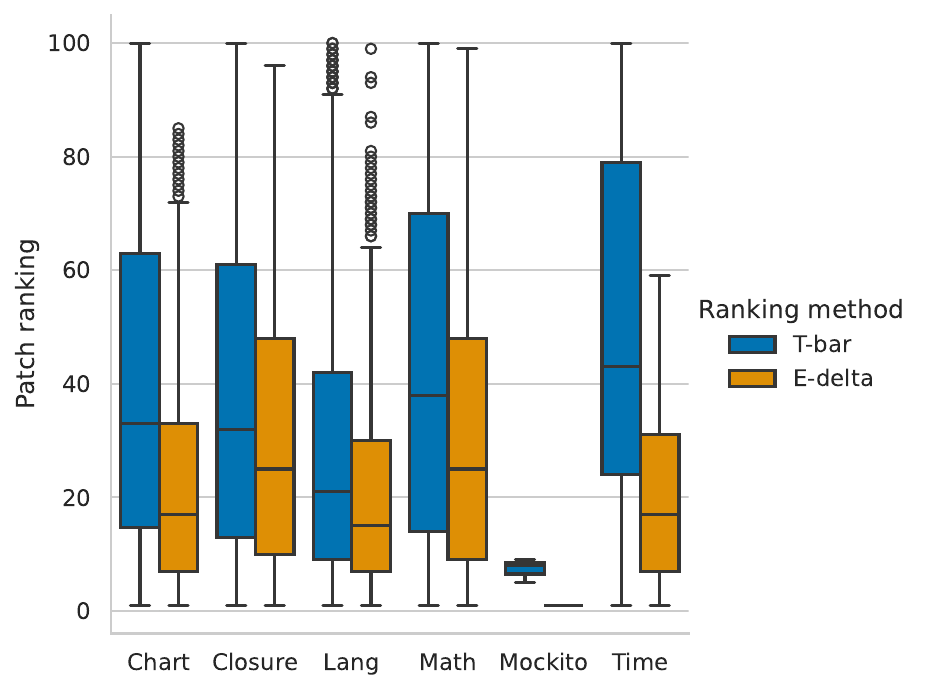}
  \caption{Entropy-delta and \tbar ranking (lower is better) of test-passing
  patches on 72 \defj bugs.}
  \label{fig:e-delta-efficiency}
\end{figure}

\begin{figure}[t]
  \centering
  \includegraphics[width=0.5\textwidth]{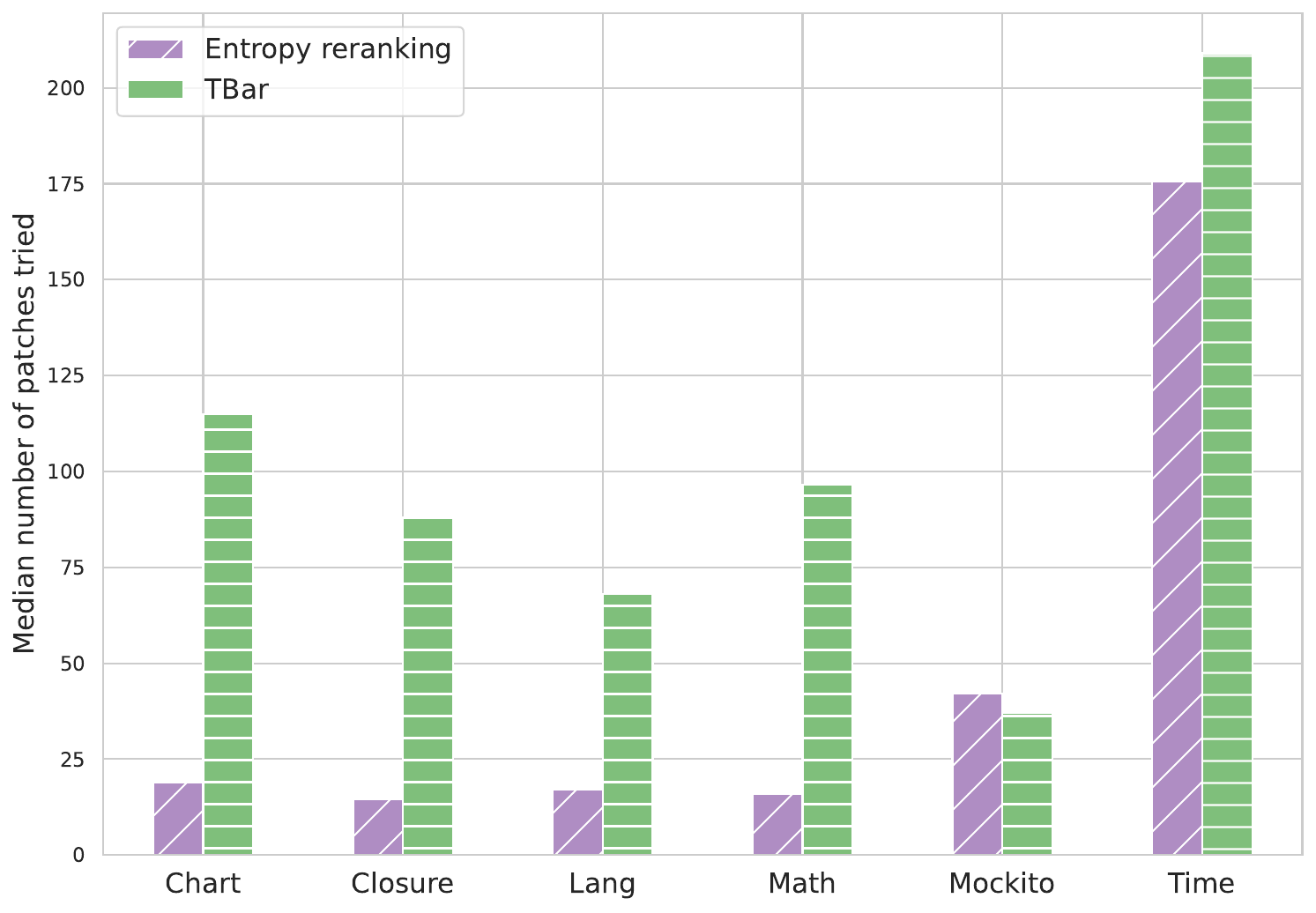}
  \caption{Median number of patches tested (lower is better) per project before succesful patch using \tbar original ranking and entropy-delta re-ranking of test-passing patches on 100 \defj bugs.}
  \label{fig:e-delta-num-patches}
\end{figure}

In this section, we discuss the observed relationship of entropy and
test-passing patches. We use entropy from \incoder, the most successful LLM in
RQ1's fault localization. We measure the impact of entropy-delta on patch
generation efficiency with two methods: (1) measuring each successful
(test-passing) patch's ranking as ranked by original \tbar and entropy-delta
re-ranked \tbar, and (2) incorporating entropy-delta into \tbar and measuring the
total number of  patches generated to pass all tests. 

We first configured \tbar to generate only 100 patches per each Defects4J bug,
assuming perfect fault localization. Of the \tbar patches we generated, 72
passed all tests contained in their bugs' respective repositories (e.g., all
tests written for project Chart). Finally, we calculated the entropy-delta score
for each patch, and the test-passing patch's original ranking according to
\tbar. As seen in Table \ref{table:patch_efficiency}, entropy-delta improves 60
out of the 72 rankings as compared to \tbar's original ranking. On average, we
observed a mean rank decrease of 24, meaning that using entropy-delta to rank
the generated \tbar patches can reduce a mean of 24 full test iterations (i.e.,
each potential patch must run through all test cases in the repository before
knowing if it is a plausible patch). Liu et
al.~\cite{efficiency} compared 16 APR techniques and found that \tbar exhibits one of the highest number of patches generated, but also the highest rate of bug fixing across \defj.
We posit that entropy-delta's efficiency improvement over \tbar significantly boosts template-based APR's overall utility.

Figure~\ref{fig:e-delta-efficiency} compares the \tbar ranking
and entropy-delta ranking. Each bar represents the rank of test-passing patches
compared to all generated patches per \defj project. A lower rank signifies a
more efficient repair process, as the repair process ends when a test-passing
patch is found. As seen in Figure~\ref{fig:e-delta-efficiency}, \tbar's original
ranking for test-passing patches is higher than entropy-delta's ranking across
all projects. Entropy-delta shows a higher disparity on ranking between test
passing and test failing patches (i.e., a lower median rank for all test-passing
patches). In particular, patches from projects Chart and Time show the largest
improvement from re-ranking patches with entropy-delta. Successful patches in
Chart and Time typically require multi-line edits, and with a wider range of
templates to choose from, entropy-delta can make a greater impact in reducing
the number of patches tested.

We then configured \tbar to use entropy-delta ranked patches directly, and
measured the total number of patches required until a successful bug fix (i.e.,
passing all tests). Figure~\ref{fig:e-delta-num-patches} shows the median number
of patches tested per project before a successful patch using \tbar original
ranking and entropy-delta re-ranking. We observe that entropy-delta re-ranking
reduces the median number of patches tested across all projects except for
Mockito. Mockito has only three single line bugs that \tbar can fix. With a smaller total number of patches
to try on a single template (e.g., 11 total possible patches for Mockito-26),
entropy-delta re-ranking does not have as large of an impact on APR efficiency. 

% Finally, we measure the total time spent on patch generation and testing. We
% our customized test-caching and patch re-ranking using entropy-delta, we
% observe a total Defects4J repair time of 42m 12s, as compared to 52m 15s 30s
% for original \tbar.

\begin{tcolorbox}
  [colback=white,colframe=black,arc=0pt,boxrule=0.5pt,title=RQ2
 Summary,boxsep=2pt,left=1pt,right=1pt,top=1pt,bottom=1pt,fonttitle=\bfseries]
 We show that entropy can be used to rank patches before going through the
 entire test-suite, thereby reducing the test overhead for template-based repair
 technique \tbar by a mean of 24 patches tested. Entropy-delta can both reduce
 the median number of patches tried before finding a fix, and consistently rank
 test patching patches higher than test-failing patches without any dependency
 on the test-suite. Entropy-delta is most useful for bugs that require
 multi-line patches.
\end{tcolorbox}

\subsection*{RQ3: How well does entropy-deltas identify correct patches?}
\begin{table*}[t]
  \centering
\caption{\small Ranking results of 1290 plausible patches per \defj project using ranking methods SBFL, Shibboleth, and entropy-delta}
\begin{tabular}{l|lrrrrrr} 
\toprule
 \textbf{Project} & \textbf{\#Patches} & \textbf{\#Correct} &
 \textbf{\#Incorrect} & \textbf{Top-N} & \textbf{SBFL} & \textbf{Shibboleth} &
 \textbf{Entropy-delta}\\
\midrule
\multirow{2}{*}{Chart} & \multirow{2}{*}{201} & \multirow{2}{*}{19} &
\multirow{2}{*}{182} & Top-1              & 3  & 11  & 10   \\
&  &  & & Top-2              & 6  & 14  & 14   \\

\midrule
\multirow{2}{*}{Closure} & \multirow{2}{*}{269} & \multirow{2}{*}{64} &
\multirow{2}{*}{205} & Top-1              & 19  & 27  & 48   \\
&  &  & & Top-2              & 38  & 47  & 58   \\

\midrule
\multirow{2}{*}{Lang} & \multirow{2}{*}{220} & \multirow{2}{*}{35} &
\multirow{2}{*}{185} & Top-1              & 1  & 14  & 20   \\
&  &  & & Top-2              & 12  & 22  & 27   \\

\midrule
\multirow{2}{*}{Math} & \multirow{2}{*}{541} & \multirow{2}{*}{67} &
\multirow{2}{*}{474} & Top-1              & 10  & 27  & 39   \\
&  &  & & Top-2              & 30  & 38  & 55   \\

\midrule
\multirow{2}{*}{Mockito} & \multirow{2}{*}{2} & \multirow{2}{*}{1} &
\multirow{2}{*}{1} & Top-1              & 0  & 1  & 1   \\
&  &  & & Top-2              & 1  & 1  & 1   \\

\midrule
\multirow{2}{*}{Time} & \multirow{2}{*}{57} & \multirow{2}{*}{11} &
\multirow{2}{*}{46} & Top-1              & 3  & 8  & 9   \\
&  &  & & Top-2              & 5  & 5  & 10   \\
\midrule
\multirow{2}{*}{Total} & \multirow{2}{*}{1290} & \multirow{2}{*}{197} &
\multirow{2}{*}{1093} & Top-1              & 36  & 85  & \textbf{127}   \\
&  &  & & Top-2              & 92  & 130  & \textbf{165}   \\
\end{tabular}
\label{table:patch_topn}
\end{table*}

% \begin{figure}
% \centering
% \includegraphics[width=0.5\textwidth]{images/sbfl_entropy_delta.pdf}
% \caption{Patch rank comparison between SBFL ranking and SBFL + entropy-delta re-ranking}
% \label{fig:rq3-rank-change}
% \end{figure}

\begin{table}[t]
  \centering
\caption{\small Classification scores of 2,147 plausible patches on \defj projects using classification methods PATCH-SIM, Panther, and entropy-delta}
\begin{tabular}{r|rrr}
\toprule
\textbf{Score}  & \textbf{PATCH-SIM}& \textbf{Panther} & \textbf{Entropy-delta}
\\
\midrule
Accuracy  & 0.388 &  0.730 & 0.735\\
Precision  & 0.245 &  0.760 &0.900\\
+ Recall & 0.711 & 0.757 &0.760\\
- Recall & 0.572 &  0.696 &0.624\\
F1  & 0.377 &  0.750 & 0.824\\
\bottomrule
\end{tabular}
\label{table:patch_classification}
\end{table}

\begin{figure}
\centering
\includegraphics[width=0.53\textwidth]{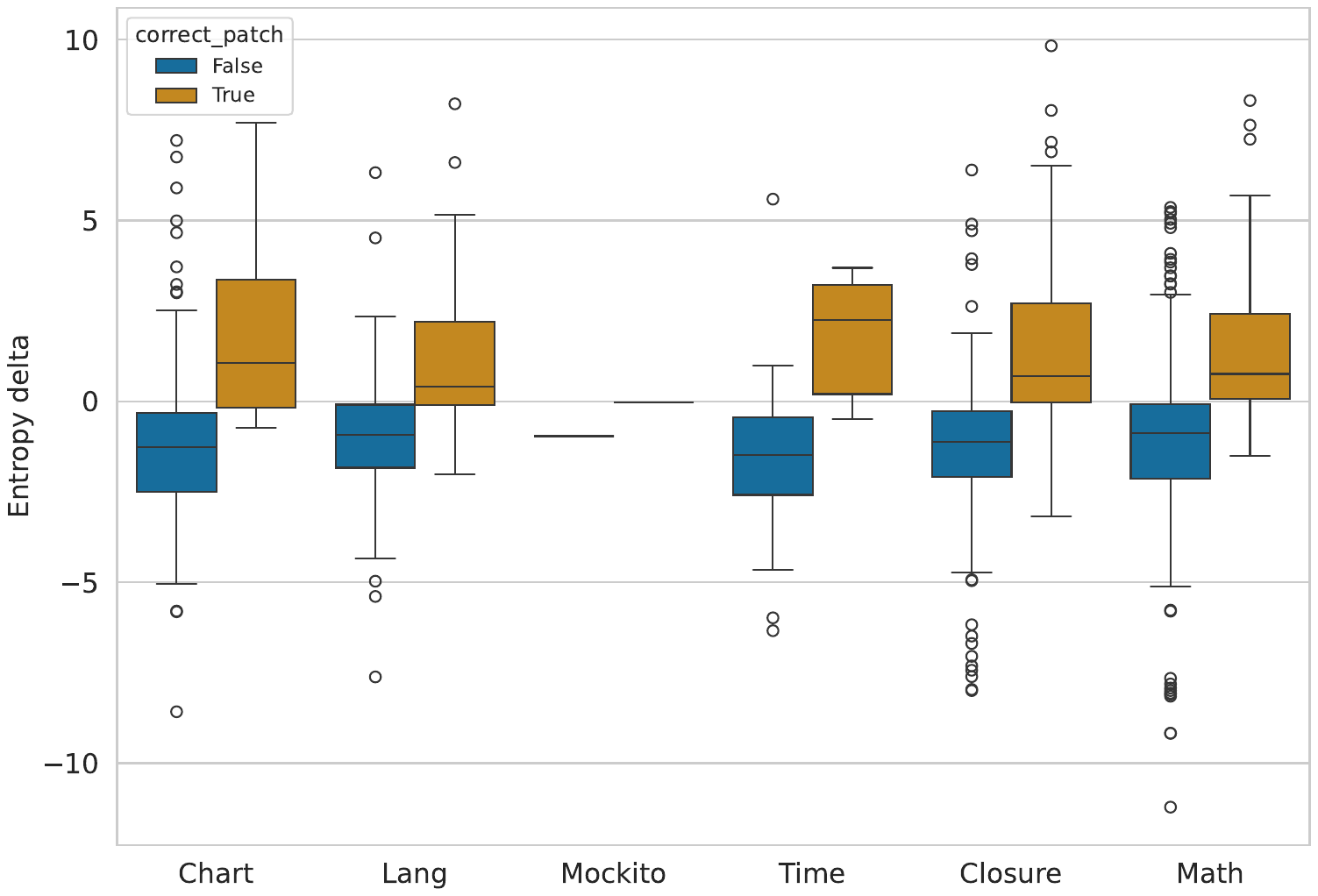}
\caption{Entropy-delta across correct and incorrect patches on \defj projects. A
higher entropy-delta signifies a less surprising patch to the LLM, and a lower
entropy (sometimes negative) entropy-delta signifies a more surprising patch to
the LLM.}
\label{fig:e-delta-patch}
\end{figure}

In RQ2, we saw that entropy-delta can improve the efficiency of patch
generation by reducing number of patches tested. However, it is important to
note that a test-passing patch is not necessarily correct. To further explore
the issue of correctness, we investigated the ability of entropy-deltas to
distinguish between correct and incorrect patches, both of which are
test-passing.

\subsubsection{Patch ranking}
We evaluate a dataset of 1,290 patches generated by 7 prior APR methods
collected by Ghanbari et al.~\cite{Shibboleth}. For each bug, the data set
includes some number of plausible (i.e., test passing) patches, where exactly
one is correct, and the rest are incorrect. We attempt to isolate the true
correct patch from the incorrect patches. We then rank each patch according to
its entropy-delta, querying the model for the entropy of the entire patch region
before and after the replacement. Table~\ref{table:patch_topn} shows the Top-1
and Top-2 results of our approach on the labeled dataset of 1,290 patches. We
see from Table~\ref{table:patch_topn} that entropy-delta outperforms both SBFL
and Shibboleth~\cite{Shibboleth} on Top-2 across all projects, and entropy-delta outperforms
Shibboleth on Top-1 across all projects but Chart (10 Top-1 as compared to
Shibboleth's 11 Top-1). Overall, we see that entropy-delta improves upon
Shibboleth by 49\% for Top-1, and 27\% for Top-2.

The difference in entropy reduction between correct and plausible but incorrect
patches is shown in greater detail in Figure~\ref{fig:e-delta-patch}. We see a
clear difference in entropy-delta across correct and incorrect patches. In
particular, the correct patches for all six projects have a median entropy-delta
value of above 0, and the incorrect patches for all six projects have a median
entropy-delta value of below 0. A correct patch tends to appear
more natural to the LLM as compared to its original buggy line.

% Similarly to
% RQ1, we experiment with how we can use entropy in addition to prior patch
% ranking tools. We observe in Figure~\ref{fig:rq3-rank-change} that entropy
% re-ranking (i.e., using entropy to rank again after a filtered SBFL list) of
% Top-5 plausible patches mostly lowered ranking. In particular, we observe that
% of our dataset of 1290 plausible patches ranked by SBFL, only 119 (9\%)
% re-ranked with entropy actually raised the correct patch ranking.  
% Our findings suggest that entropy-delta is a powerful tool for highlighting the
% correct patch among a series of plausible options. 

% \ruben{Not sure if the focus on SBFL is worth it here. Figure 7 shows that we can improve the ranking which is already seen before. How many Top-1 and Top-2 would SBFL+entropy have? Could have we have done something similar with Shibboleth?}
% \aidan{I did not do a shibboleth rerank using entropy, Fig 7 was from a long time ago when Sophia was still working on this project. 
% Maybe it is not worth showing anymore. I commented it out for now.}

\subsubsection{Patch classification}

Table~\ref{table:patch_classification} shows our classification results on a
labeled dataset of 2,147 plausible patches curated by Tian et al.~\cite{Panther}
for classifying patches as correct or incorrect. Entropy-delta 
improves upon the accuracy score of PATCH-SIM~\cite{patch-sim} and Panther~\cite{Panther}, but only slightly
improves +recall score over both PATCH-SIM and Panther. For -recall,
entropy-delta performs better than PATCH-SIM by 9\%, but performs worse than
Panther by 10\%. Entropy-delta slightly improves accuracy over Panther by 0.6\%,
and 89\% over PATCH-SIM. Entropy-delta improves precision over Panther by 18\%,
and PATCH-SIM by 267\%. Finally, entropy-delta performs better than both
PATCH-SIM and Panther on F1 score, by 118\% and 10\% respectively. As compared to the state-of-the-art, entropy improves classification performance on true positives more than true negatives. 

%\ruben{Can we justify why do we have worse -recall?} \aidan{I really don't know why, just wrote something to address it.}

Our analysis focused on comparing the degree of entropy reduction between true
correct patches and plausible test-passing patches. As shown in
Table~\ref{table:patch_topn}, Table~\ref{table:patch_classification}, and
Figure~\ref{fig:e-delta-patch}, our results suggest that correct patches
tend to lower entropy (i.e., increase naturalness) more than incorrect patches. 
Specifically, entropy-delta ranks 49\% more correct patches in the Top-1
than the state-of-the-art patch ranker Shibboleth, and entropy-delta can
classify correct patches with an 18\% higher precision than the state-of-the-art
patch classifier Panther. These findings suggest that entropy-deltas can be a
valuable heuristic for distinguishing between correct and incorrect patches.

\begin{tcolorbox}
  [colback=white,colframe=black,arc=0pt,boxrule=0.5pt,title=RQ3
    Summary,boxsep=2pt,left=1pt,right=1pt,top=1pt,bottom=1pt,fonttitle=\bfseries]
    The entropy-delta from an LLM distinguishes between correct and plausible
    (test-passing but incorrect) patches with
    higher precision and accuracy than state-of-the-art patch disambiguation tools.
\end{tcolorbox}

\section{Related Work}
\label{sec:related}

We discuss in the following sections the most recent advances in LLM for code,
fault localization, and patch ranking.

\subsection{LLM for code}
Language models have been used for code generation, bug detection, and patch
generation. Recent language models finetune on code as training data and can
perform code completion~\cite{desai2016program, inCoder}, and generate code
based on natural language~\cite{raychev2014code} with impressive results. Large
Language Models (LLMs), such as Codex~\cite{Codex}, GPT-Neo~\cite{Neox}, and
Llama-2~\cite{Llama} have raised performance on these tasks by using more
trainable parameters and training data. Ray et al.~\cite{NAT} study the
relationship between bugginess and LLM-entropy. Ray et al. empirically showed
that n-gram models trained over a large corpus of code will find buggy
statements more surprising, as indicated by a high entropy score. Kolak et
al.~\cite{LLM-PATCH} revisit the question of naturalness (i.e., the
human-readability) of patches in the era of large language models. Kolak et al.
experimented with models ranging from [160M to 12B] parameters, and measured the
similarly between LLM generated patches and developer written patches. Their
results show that larger models tend to generate test-passing lines at a higher
rate. Additionally, LLM generated patches tend to be more similar to the
human-written patch as model size increases. Xia et al.~\cite{xia2023automated}
directly applied LLMs for APRs and found that LLMs can suggest multi-line fixes
with higher accuracy than state of the art APR tools. Our study performs an
empirical evaluation of how code naturalness (i.e., entropy) can improve prior
APR tools across three different stages of automated program repair: fault
localization, patch generation, and patch ranking.

\subsection{Fault localization}
Prior fault localization tools use test output information, code semantics, and
naturalness of code to achieve a high degree of confidence on bug detection.
Spectrum-based Fault Localization (SBFL)~\cite{abreu2006evaluation,
abreu2007accuracy} uses a ratio of passed and failed tests covering each line of
code to calculate its suspiciousness score, in which a higher suspiciousness
signifies a higher probability of being faulty. Recent advances in deep learning
created a spur of research on using graph neural networks (GNNs)~\cite{GNN} for
fault localization. GRACE~\cite{Grace}, DeepFL~\cite{DeepFL}, and
DEAR~\cite{Dear} encode the code AST and test coverage as graph representations
before training deep learning models for fault localization.
TransferFL~\cite{TransferFL} combined semantic features of code and the
transferred knowledge from open-source code data to improve the accuracy of
prior deep learning fault localization tools. LLMAO~\cite{Llmao} finetuned
a light-weight bidirectional layer on top of code-tuned LLMs to show that LLMs
can detect both bugs and security vulnerabilities without the use of test cases.
Our work builds on top of the top-performing prior fault localization tools and
show that entropy can be used as a light weight re-ranking tool that improves
fault localization scores without a dependency on test cases.

\subsection{Patch correctness}
Similarly to prior fault localization tools, prior patch disambiguation tools
leverage test output information and code information (both code syntax and code
semantics) for ranking or classifying patches. Qi et al.~\cite{qi2015analysis} analyzed the reported
bugs of three generate-and-validate APR tools: GenProg~\cite{genprog},
RSRepair~\cite{rsrepair}, and AE~\cite{ae} systems, to find that producing
correct results on a validation test suite is not enough to ensure patch
correctness. Smith et al.~\cite{CURE} performed
an experiment that interrogates whether or not automatically generated patches
are prone to overfitting to their test suite. Borrowing the concept of training
and test sets from machine learning, they found that automated program repair
(APR) typically used the same test-suite for both ``training" (generating the
patch), and ``testing" (validation). Smith et al. found that both the coverage
rate of the test-suite, as well as the assignment of test/train sets between the
two suites, impact the degree of overfitting in repair. To counteract the
overfitting problem, Ye et al.~\cite{ye_patches} proposed ODS (Overfitting
Detection System), a novel system to statically classify overfitting patches.
Xiong et al.~\cite{patch-sim} generated both execution traces of patched
programs and new tests to assess the correctness of patches. Ghanbari et
al.~\cite{Shibboleth} used both the syntactic and semantic similarity between
original code and proposed patch, and code coverage of passing tests to rank
patches. Shibboleth~\cite{Shibboleth} was able to rank the correct patch in
Top-1 and Top-2 positions in 66\% of their curated dataset. Tian et
al.~\cite{Panther} proposed machine learning predictor with BERT
transformer-based learned embeddings for patch classification. Tian et al. found
that learned embeddings of code fragments with BERT~\cite{BERT},
CC2Vec~\cite{cc2vec}, and Doc2Vec~\cite{doc2vec} yield similarity scores that,
given a buggy code, substantially differ between correctly-patched code and
incorrectly-patched one.

% Yang et al.~\cite{yang_patch} performed an empirical evaluation of 9 prior
% patch disambiguation tools and observed that static techniques incorrect
% assumed that correct patches are more similar to buggy code than incorrect
% plausible patches. 

The most relevant work to our study of patch correctness is Yang et al.~\cite{yang2023large}. Yang et
al.~\cite{yang2023large} found that state-of-the-art learning-based techniques
suffered from the dataset overfitting problem, and that naturalness-based
techniques outperformed traditional static techniques, in particular
Patch-Sim~\cite{patch-sim}. Our work uses 2,147 plausible patches collected in
2023 (past the LLM training data cutoff of all our chosen LLMs), which lowers
the risk of LLM training data leakage. Our work performs an empirical study on
entropy against the most recent state-of-the-art patch disambiguation techniques
Panther~\cite{Panther} and Shibboleth~\cite{Shibboleth}, on top of
Patch-Sim~\cite{patch-sim}. Motivated by Liu et al.~\cite{efficiency}, our work
is the first to use LLM entropy on plausible patches before undergoing testing
to achieve more efficient APR on prior template-based techniques. Finally, we
introduce a new naturalness measurement for patches, entropy-delta, which
achieves state-of-the-art results for plausible patch disambiguation without
depending on the test-suites of buggy programs, which lowers the risk of dataset
overfitting.

\section{Threats}
\label{sec:threats}
\textbf{External validity.}
A threat to the external validity of our study is the potential selection bias
of our three selected LLMs. We chose a representative set of LLMs with a range
of trainable parameters. We chose the three based on their infill ability and 
built-in bidirectional attention mechanism. Much larger LLMs ($> 20$
billion parameters) might have a stronger ability to reason over faulty code
lines and patches, but require much larger computation power and time for
entropy calculation. Another threat to external validity is our usage of \defj data
throughout our empirical evaluation. We chose \defj for our target bugs for
fault localization and patch disambiguation due to the data available and the
aim to compare against related work in APR. Data leakage of \defj as training
data for our selected LLMs is possible. We mitigate this risk by (1) using
entropy in combination with prior APR techniques instead of direct LLM prompting
for patch generation, and (2) applying entropy-delta on untested or plausible
patches that are not available online (i.e., recently generated and not used as
an official patch for bug fixing).

\textbf{Internal validity.}
An internal validity is the manual labeling of plausible patches. We used
manually labeled data released by prior works~\cite{Panther, Shibboleth}, in
which the authors followed clear and reproducible decision criteria. Although
mistakes could still be made on which plausible patches are correct or
incorrect, we use the same labeling for all prior tools studied as well as
entropy to create a standardized baseline on patch classification.

\textbf{Construct validity.}
One source of construct validity is the measurements we chose for our empirical
evaluation. We used Top-N as a ranking measurement for both bugs and patches,
following prior APR work. To overcome some limitations of Top-N, we also use
multiple patch classification measurements (accuracy, precision, recall, and F1)
on a separate set of labeled patch data to strengthen generalizability.

\section{Conclusion}
\label{sec:conclusion}
In this work, we propose the use of ``unnaturalness'' of code for automated
program repair through the measurement of entropy generated by code-tuned LLMs.
We also introduce the term entropy-delta, which measures the difference in
entropy between a proposed code insert (i.e., a patch) and the original code.
Using three LLMs and three prior fault localization tools, we show that entropy
can improve Top-5, 3, and 1 scores after re-ranking the first 6 potential bug
localization. We use entropy-delta on untested patches to save an average of 24
test runs per bug for the template-based APR technique TBar. We show that
entropy-delta can improve upon state-of-the-art patch ranking by 49\% for Top-1,
and classify plausible patches with a 18\% higher precision. Our results
indicate that LLMs can be a powerful addition to state-of-the-art APR tools
without the dependency on tests, and the usage of LLM code-generation. The
reduction in both test suites and LLM code-generation results in the reduction
in model over-fitting and training data leakage.

\bibliographystyle{IEEEtran}
\bibliography{ref}

\end{document}